
\input phyzzx
\baselineskip=23pt plus 1pt minus 1pt
\hoffset=0mm
\hsize=160mm
\voffset=0mm
\vsize=230mm
\tolerance=10000

\def\Vec#1{
   \hbox to10mm{{\sl #1} \hfill}\hskip-9.88mm
   \hbox to10mm{{\sl #1} \hfill}\hskip-9.88mm
   \hbox to10mm{{\sl #1} \hfill}\hskip-9.88mm
   {\sl #1}}
\def\bx#1{\setbox1\hbox{#1} \dimen1=\dp1\advance\dimen1 3.5pt%
   \hskip1pt\lower\dimen1
   \hbox{\vrule\vbox{\hrule\vskip3pt%
   \hbox{\hskip3pt#1\hskip3pt}%
   \vskip3pt\hrule}\vrule}\hskip3pt}

\def\ie{{\it i.e.}}
\def\refmk#1{$^{#1)}$}
\def\figmk#1{\medskip \centerline{\bx{Fig.~#1}} \medskip}
\def\refs#1{\item{#1 )~}}

\REF\AriMat{}
\REF\Bengtsson{}
\REF\HolAb{}
\REF\NazDob{}
\REF\SakKis{}
\REF\Gutz{}
\REF\BarDav{}
\REF\Wintgen{}
\REF\Nextpaper{}

\FIG\POC{}
\FIG\FTLA{}
\FIG\TRM{}

\rightline{\hbox to22mm{KUNS 1196\hfill}}
\rightline{\hbox to22mm{April 1993\hfill}}
\bigskip

\centerline{\fourteenbf
  Supershell Effect and Stability of Classical Periodic Orbits}
\centerline{\fourteenbf
  in Reflection-Asymmetric Superdeformed Oscillator}
\medskip
\centerline{\fourteenrm
  Ken-ichiro A{\twelverm RITA}}
\medskip
\centerline{\it Department of Physics, Faculty of Science,}
\centerline{\it Kyoto University, Kyoto 606-01}
\medskip

\centerline{\bf Abstract}
\midinsert\narrower
{
\tenpoint
\baselineskip=15pt
\noindent
A semiclassical analysis is made of the origin of an undulating
pattern in the smoothed level density for a reflection-asymmetric
superdeformed oscillator potential.
It is suggested that, when the octupole-type deformation increases,
an interference effect between two families of periodic orbit
with the ratio of periods approximately 2:1 becomes stronger
and thus a pronounced ``supershell'' structure appears.
}
\endinsert
\bigskip


  The quantum-energy spectrum
in the axially-symmetric oscillator potential
with frequency ratio $\omega_\perp/\omega_z$=2
(called ``superdeformed'' oscillator)
is known to have ``supershell'' structure
(\ie, modulation with periodicity $2\hbar\omega_{\rm sh}$
in the oscillating level density,
$\omega_{\rm sh}$ being the basic frequency
of the superdeformed oscillator).
We have indicated in our previous paper\refmk{\AriMat} that
the supershell effect is significantly enhanced
when octupole $(Y_{30})$ deformation is added
to the 2:1 deformed harmonic-oscillator potential,
and suggested that this enhancement might be responsible
for the odd-even effect
(with respect to the shell quantum number $N_{\rm sh}$)
in stability of superdeformed states against octupole deformed shape,
discussed by Bengtsson et al.,\refmk{\Bengtsson}
H\"oller and {\AA}berg,\refmk{\HolAb}
and Nazarewicz and Dobaczewski.\refmk{\NazDob}
The single-particle Hamiltonian we used in our analysis is
$$
h={~\Vec{p}^2\over 2M}
  +\sum_{i=x,y,z}{M\omega_i^2x_i^2\over 2}
  -\lambda_{3K}M\omega_0^2\left(r^2\,Y_{3K}(\Omega)\right)'',
\eqn\Hamil
$$
where $\omega_x$=$\omega_y$=$2\omega_z\equiv 2\omega_{\rm sh}$
and $\omega_0^3$=$\omega_x\omega_y\omega_z$.
The double primes denote that the variables in parenthesis are
defined in terms of the doubly-stretched coordinates\refmk{\SakKis}
$x_i''=(\omega_i/\omega_0)\,x_i$.
In the following, we limit to the case $K=0$.
When the variables are scale-transformed to dimensionless ones,
the Hamiltonian \Hamil~ is rewritten as
$$
h={~\Vec{p}^2\over 2}
  +\left({~r^2\over 2}-\lambda_{30}\,r^2\,Y_{30}\,(\theta)\right)'',
\eqn\Hamiltn
$$
where $x''$=$2x$, $y''$=$2y$ and $z''$=$z$.

  As is well known, the quantum level density
$g\,(E)=\sum_n \delta\,(E-E_n)$
is expressed in semiclassical theory as\refmk{\Gutz}
$$
g\,(E)\simeq {\bar g}\,(E)
  +\sum_\gamma A_\gamma(E)
  \cos\left(S_\gamma(E)/\hbar-\hbox{(phases)}_\gamma\right),
\eqn\Trformula
$$
where ${\bar g}$ is an average level density
corresponding to the Thomas-Fermi approximation,
and the second term in the r.h.s. represents
the oscillatory contributions from periodic orbits,
$S_\gamma$ is the action integral
$\oint_\gamma\Vec{p}\cdot d\Vec{q}$,
and the amplitude factor $A_\gamma$ is mainly related to
the stability of the orbit $\gamma$.
When one is interested in an undulating pattern in $g\,(E)$
smoothed to a finite resolution $\delta E$ (\ie, shell structure),
it is sufficient to only consider short periodic orbits
with the periods $T_\gamma<2\pi\hbar/\delta E$.
The supershell structure is expected to arise
from interference effects between orbits with different periods
$T_\gamma$.
The short periodic orbits are calculated by Monodromy
Method\refmk{\BarDav} and shown in Fig.~\POC.

\figmk{\POC}

\noindent
For the Hamiltonian system under consideration
whose phase space is constructed with both regular and chaotic
regions, evaluation of $A_\gamma$ in eq.~\Trformula~ is not always
easy because the stationary-phase approximation breaks down near
resonances which take places rather frequently in the regular
regions.
Fortunately, however, by virtue of the scaling property,
$h\,(\alpha\Vec{p},\alpha\Vec{q})=\alpha^2h\,(\Vec{p},\Vec{q})$,
we can use the Fourier-transformation techniques\refmk{\Wintgen}
and extract informations about classical periodic orbits
from quantum energy spectrum.
The scaling rules for variables in eq.~\Trformula~
are\refmk{\Nextpaper}
$$
\eqalign{
  {\bar g}\,(E)&=E^2{\bar g}\,(1), \cr
  S_\gamma(E)&=E\,T_\gamma, \cr
  A_\gamma(E)&=E^{k_\gamma}A_\gamma(1) \hskip8mm
    \cases{k_\gamma=0  &for isolated orbits, \cr
           k_\gamma={1\over 2}  &otherwise. \cr} \cr
}
\eqn\Scaling
$$
The last equality is obtained under the stationary-phase
approximation.
Using these relations, it is easy to see that
the Fourier transform of eq.~\Trformula~
multiplied by an appropriate weighting factor $E^{-k}$
will exhibit peaks at the periods $T_\gamma$ of classical periodic
orbits and the heights of the peaks represent the strengths of their
contributions.
In Fig.~\FTLA, we show the power spectrum $P\,(s)$
for several values of $\lambda_{30}$,
taking $k={1\over 2}$ appropriate to non-isolated orbits;
$$
P\,(s)=\left|~\sum_n{e^{isE_n}\over\sqrt{E_n}}~\right|~.
\eqn\Power
$$

\figmk{\FTLA}

\noindent
We see nice correspondence between peak locations of $P\,(s)$
and periods of classical periodic orbits.
The most important observation is that
relative intensity between peaks
at $s\approx\pi$ and $s\approx2\pi$ changes
as the octupole deformation parameter $\lambda_{30}$ increases.
This result indicates that the enhancement of the supershell effect
(shown in Fig.~6 of Ref.~1) ) may be explained as
due to the growth of the interference effect between
classical periodic orbits with periods $T\approx\pi$
and those with $T\approx2\pi$.

  To understand the cause of the change
in relative intensity mentioned above,
let us investigate properties of the classical periodic orbits.
Calculating periodic orbits by the Monodromy Method,\refmk{\BarDav}
we obtain stability matrices $M_\gamma$ for the periodic orbits
$\gamma$.
They are linearized Poincar\'e maps at the periodic orbits
defined as
$$ \left({\delta\Vec{p}\,(T_\gamma) \atop \delta\Vec{q}\,
  (T_\gamma)}\right)=M_\gamma\left({\delta\Vec{p}\,(0) \atop
  \delta\Vec{q}\,(0)}\right)+{\cal O}\,(\delta^2),
\eqn\Smtrx
$$
where $(\delta\Vec{p},\delta\Vec{q})$ represent deviations
from the periodic orbits $\gamma$ in phase space.
These six-dimensional matrices $M_\gamma$ are real and symplectic,
so that eigenvalues of each $M_\gamma$ appear in pairs
$\pm(e^{\alpha},e^{-\alpha})$,
$\alpha$ being real or pure imaginary.
As the Hamiltonian \Hamiltn~ is axially-symmetric,
classical orbits are usually non-isolated.
For such orbits, each stability matrix has 4 unit eigenvalues.
Values of $\Tr M$ written in Fig.~\POC~
are sums of the remaining 2 eigenvalues
which determine stabilities of the periodic orbits;
$\alpha$=$iv$ is pure imaginary and $|\Tr M|=|2\cos v|\leq 2$
when the orbit is stable,
while $\alpha$=$u$ is real and $|\Tr M|=|\pm 2\cosh u|>2$
when the orbit is unstable.
Under the stationary-phase approximation,
the amplitude factors $A_\gamma$ in eq.~\Trformula~
are inversely proportional to $\sqrt{|\Tr M_\gamma-2|}$.
Fig.~\TRM~ shows values of $\Tr M$ for relevant orbits
calculated as functions of the octupole-deformation parameter
$\lambda_{30}$.

\figmk{\TRM}

\noindent
{}From this figure, we see that
the orbits with $T\approx \pi$ are always stable
and their values of $\Tr M$ approach to 2 with increasing
$\lambda_{30}$, while the orbit B (C,C') with $T\approx 2\pi$
become unstable (more unstable).
Therefore, we can expect that
the contributions from orbits with $T\approx\pi$ increase
when $\lambda_{30}$ becomes large,
while those from orbits with $T\approx 2\pi$ decrease.
This result suggests that the enhancement of the supershell effect
stems from the difference of the stability against octupole
deformation between these two families of periodic orbit.

  A more detailed analysis of the supershell structure
in reflection-asymmetric superdeformed oscillator potentials
will be given in a forthcoming full-length paper.\refmk{\Nextpaper}
The author thanks Prof. Matsuyanagi for carefully reading
the manuscript.

\bigskip
\bigskip

\centerline{\bf References}
\medskip
\itemsize=25pt
\refs{ 1}
  K. Arita and K. Matsuyanagi,
  Prog. Theor. Phys. {\bf 89} (1993), 389.
\refs{ 2}
  T. Bengtsson, M.E. Faber, G. Leander, P. M\"oller, M. Ploszajczak,
  I. Ragnarsson and S. {\AA}berg,
  Physica Scripta {\bf 24} (1981), 200.
\refs{ 3}
  J. H\"oller and S. {\AA}berg,
  Z. Phys. {\bf A336} (1990), 363.
\refs{ 4}
  W. Nazarewicz and J. Dobaczewski,
  Phys. Rev. Lett. {\bf 68} (1992), 154.
\refs{ 5}
  H. Sakamoto and T. Kishimoto,
  Nucl. Phys. {\bf A501} (1989), 205.
\refs{ 6}
  M.C. Gutzwiller,
  J. Math. Phys. {\bf 8} (1967), 1979;
  {\it ibid.} {\bf 12} (1971), 343.
\refs{ 7}
  M. Baranger, K.T.R. Davies and J.H. Mahoney,
  Ann. of Phys. {\bf 186} (1988), 95.
\refs{ 8}
  See, for instance,
  H. Friedrich and D. Wintgen,
  Phys. Rep. {\bf 183} (1989), 37.
\refs{ 9}
  K. Arita and K. Matsuyanagi,
  in preparation.

\endpage
\nopagenumbers

\hbox{}\vfill\noindent Fig.~1.~~
  Short periodic orbits for the Hamiltonian \Hamiltn~
  with $\lambda_{30}=0.4$.
  Upper part: Planar orbits in the plane containing the symmetric
  axis $z$.
  Lower part: A circular orbit in the plane perpendicular to the
  symmetry axis (A') and a three-dimensional orbit (C').
  Their projections on the $(x,y)$ plane and on the $(z,y)$ plane
  are shown.
\endpage

\hbox{}\vfill\noindent Fig.~2.~~
  Power spectra $P\,(s)$ defined by eq.~\Power~
  for $\lambda_{30}=0.2\sim 0.4$.
  The summation is taken up to $n$=200.
  Arrows indicate periods of the classical periodic orbits
  (see Fig.~\POC) and their repetitions.
\endpage

\hbox{}\vfill\noindent Fig.~3.~~
  Traces of the stability matrices $M$ for the periodic orbits
  shown in Fig.~\POC~ (see text for their definitions).
  For the isolated orbit A',
  the stability matrix $M$ has 2 unit eigenvalues
  and the remaining 4 eigenvalues appear in two pairs
  $(e^{\alpha_a},e^{-\alpha_a})$ and $(e^{\alpha_b},e^{-\alpha_b})$.
  Thus, A'$_a$ and A'$_b$ denote the traces of these pairs,
  respectively.
\endpage

\end